\documentclass[12pt,epsf,epsfig,psfig]{article}
\usepackage{graphicx}
\usepackage{amsmath}
\usepackage{amssymb}
\usepackage{mathrsfs}
\usepackage{graphicx}
\usepackage{epsfig}
\usepackage{cite}
\begin{document}

\vspace{4cm}
\begin{center}
{\Large\bf{The Angular Correlations in Top Quark Decays in Standard Model Extensions}}\\

\vspace{1cm}
{\bf  S. Batebi, S. M. Etesami$^{\dagger,\ddagger}$ }
 {\bf and M. Mohammadi Najafabadi$^{\dagger}$ }  \\
\vspace{0.5cm}
{\sl ${\ ^{\dagger}}$  School of Particles and Accelerators, \\
Institute for Research in Fundamental Sciences (IPM) \\
P.O. Box 19395-5531, Tehran, Iran}\\
and \\
{\sl ${\ ^{\ddagger}}$  Physics Department, \\
Isfahan University of Technology (IUT) \\
P.O. Box 11365-9161, Isfahan, Iran}\\

\vspace{2cm}
 \textbf{Abstract}\\
 \end{center}

The CMS collaboration at the CERN-LHC have searched for the $t$-channel
single top quark production using the spin correlation of the $t$-channel. The signal extraction
and cross section measurement relies on the angular distribution of the charged
lepton in the top quark decays, the angle between the charged lepton momentum
and top spin in the top rest frame. The behavior of the angular distribution is
a distinct slope for the t-channel single top (signal) while it is flat for the backgrounds. In this letter
we investigate the contributions which this spin correlation may receive
from a two-Higgs doublet model, a top-color assisted technicolor (TC2) and the
noncommutative extension of the Standard Model.
\\
PACS number(s): 14.65.Ha, 12.60.-i, 11.10.Nx, 12.60.Fr, 12.60.Nz

\newpage

\section{Introduction}

At the LHC, top quarks are produced via two independent
mechanisms: The dominant production mechanism is via strong interactions
where top quarks are produced in pair ($gg\rightarrow t\bar{t},q\bar{q}\rightarrow t\bar{t}$) 
and electroweak or single top quark production which occurs through three different processes: $t$-channel 
(the involved $W$-boson is space-like, $ub\rightarrow dt$), $s$-channel
(the involved  $W$-boson is time-like, $u\bar{d}\rightarrow \bar{b}t$)
and $tW$-channel (the involved $W$-boson is real, $gb\rightarrow W^{-}t$) \cite{werner}.

The NLO top pair production cross section at the LHC with the center-of-mass energy of 7 TeV is 163 pb \cite{ttbar}.
The $t$-channel single top quark with the cross section of around 62 pb
is the largest source of single top at the LHC.

Because single top quarks are produced through the electro-weak interactions, they are highly polarized.
It has been shown that there is a basis in which
the top quarks are almost $100\%$ polarized. In $t$-channel the top quark is
produced with the spin direction fully aligned with the momentum of $d$-type
quark (spectator quark) in the final state \cite{mahlon1}.
Since the top quark decays via the weak interaction, its spin is analyzed by the angular distribution
of its decay products. The differential decay rates in the rest frame of the decaying top may be parameterized
as
\begin{eqnarray}\label{angle}
\frac{1}{\Gamma}\frac{d\Gamma}{d\cos\theta_{i}}=\frac{1}{2}(1+\alpha_{i}\cos\theta_{i})
\end{eqnarray}
where $\theta_{i}$ is the angle between the chosen spin axis and the direction of motion of the $ith$
decay product, $i =l,\nu_{l},b$. In the SM, the LO values for spin analyzing powers
are: $\alpha_{l} = 1, \alpha_{b} = -0.387, \alpha_{\nu} = -0.33$ \cite{mahlon1}.
The NLO QCD corrections to the spin analyzing power of charged lepton is around $0.1\%$ while
it is $5\%$ for the $b$-quark \cite{nlo}.
The most sensitive spin analyzer in the semi-leptonic top decay is the charged lepton with $\alpha_{l}=1$.
The CMS collaboration at CERN performed a search for the $t$-channel single top
using the introduced angular distribution of the charged lepton. In \cite{pascms},
it has been shown that the distribution of the angle between the
charged lepton momentum and the orientation of the light jet (spectator jet) is
a distinct slope for single top events while it is flat for backgrounds ($W+jets, t\bar{t},...$).
Therefore, the cross section of the $t$-channel is determined by using the above
charged lepton angular distribution. In \cite{pascms}, \cite{patrick}, the 'robustness' of
the charged lepton angular distribution against the variations of parton distribution functions,
the center-of-mass energy of the colliding protons, factorization scale, jet energy scale and some
other factors has been examined. It is worth mentioning that the charged lepton angular distribution
is robust against major detector systematics such as jet energy scale (JES), b-tagging uncertainty and missing
transverse energy (MET) uncertainty \cite{pascms}.
Therefore, the angular distribution is stable enough to be
used for $t$-channel signal extraction from backgrounds and measurement of its cross section.

In this paper our main concern is to investigate whether the spin analyzing power of the charged lepton $(\alpha_{l})$
and the b-quark $(\alpha_{b})$ receive large contributions from the beyond Standard Models such as
two-Higgs doublet model (2HDM) and top-color assisted technicolor (TC2) or they do not change considerably. 
Section is dedicated to computation of contributions from the beyond Standard Models such as
2HDM, TC2 and noncommutative extension of the SM to the spin analyzing powers. Section 3 concludes the paper.

\section{Spin Analyzers and BSM Effects}

A model independent approach in search for the new physics in the $Wtb$
vertex could be made by defining the effective Lagrangian \cite{efflag}. The
amplitude $\mathcal{M}(t(p_{t})\rightarrow b(p_{b})+W^{+}(q))$, where all particles
are on-shell, can be written as:
\begin{eqnarray}
\mathcal{M}=-\frac{g}{\sqrt{2}}\bar{u}(p_{b})[(V_{tb}P_{L}+V_{R}P_{R})\gamma^{\mu}-\frac{i\sigma^{\mu\nu}q_{\nu}}{m_{W}}(g_{L}P_{L}+g_{R}P_{R})]
u(p_{t})\epsilon^{*}_{\mu}(q)
\end{eqnarray}
where $P_{R,L}=\frac{1\pm \gamma_{5}}{2}$, $V_{tb}$ is the Cabbibo-Kobayashi-Maskawa (CKM)
matrix element. $V_{R},g_{L,R}$ are anomalous couplings contributions.
The $95\%$ C.L. upper and lower limits on the above anomalous couplings from the measured branching ratio
of the $b\rightarrow s+\gamma$ are: $-0.0007<V_{R}<0.0025,-0.0015<g_{L}<0.0004,-0.15<g_{R}<0.57$ \cite{bohdan}.
The dependence of spin analyzing powers ($\alpha_{l,b,\nu}$) on the anomalous couplings has been
studied in \cite{mojtaba1,aguilar}. In \cite{mojtaba1}, we have calculated the spin analyzing powers
as functions of anomalous couplings with the assumption that these couplings are real.
The explicit analytical relations of the $\alpha_{i}$ could be found
in \cite{aguilar} with taking the couplings in general complex.
The spin analyzing powers are mostly sensitive to $g_{R}$ and  there is no significant
dependence on the other anomalous couplings. For example, the spin analyzing powers of the
$b-$quark has the following form \cite{mojtaba1,aguilar}:
\begin{eqnarray}
\alpha_{b}=-\frac{(r^{2}-2)(V_{tb}^{2}-V_{R}^{2})+(2r^{2}-1)(g_{L}^{2}-g_{R}^{2})+2r(V_{R}g_{L}-V_{tb}g_{R})}
{(r^{2}+2)(V_{tb}^{2}+V_{R}^{2})+(2r^{2}+1)(g_{L}^{2}+g_{R}^{2})+6r(V_{R}g_{L}+V_{tb}g_{R})}
\end{eqnarray}
where $r=\frac{m_{t}}{m_{W}}$. The charged lepton ($b$-quark) spin analyzing power, $\alpha_{l} (\alpha_{b})$,
can change up to $5\% (\approx 60\%)$ when $g_{R}$ varies in the interval of $[-0.2,0.2]$ \cite{mojtaba1,aguilar}.

Within the frameworks of 2HDM and TC2, we compute the contributions of
these two beyond SM to anomalous couplings $V_{R},g_{L},g_{R}$. The program for
this calculation was taken from the authors of \cite{martin11}.
In the following we will see that $\alpha_{l} (\alpha_{b})$
can deviate from the SM value at most $0.02\% (3\%)$.

\subsection{Two-Higgs doublet model}
\begin{figure}
  \includegraphics[width=7.5cm,height=6cm]{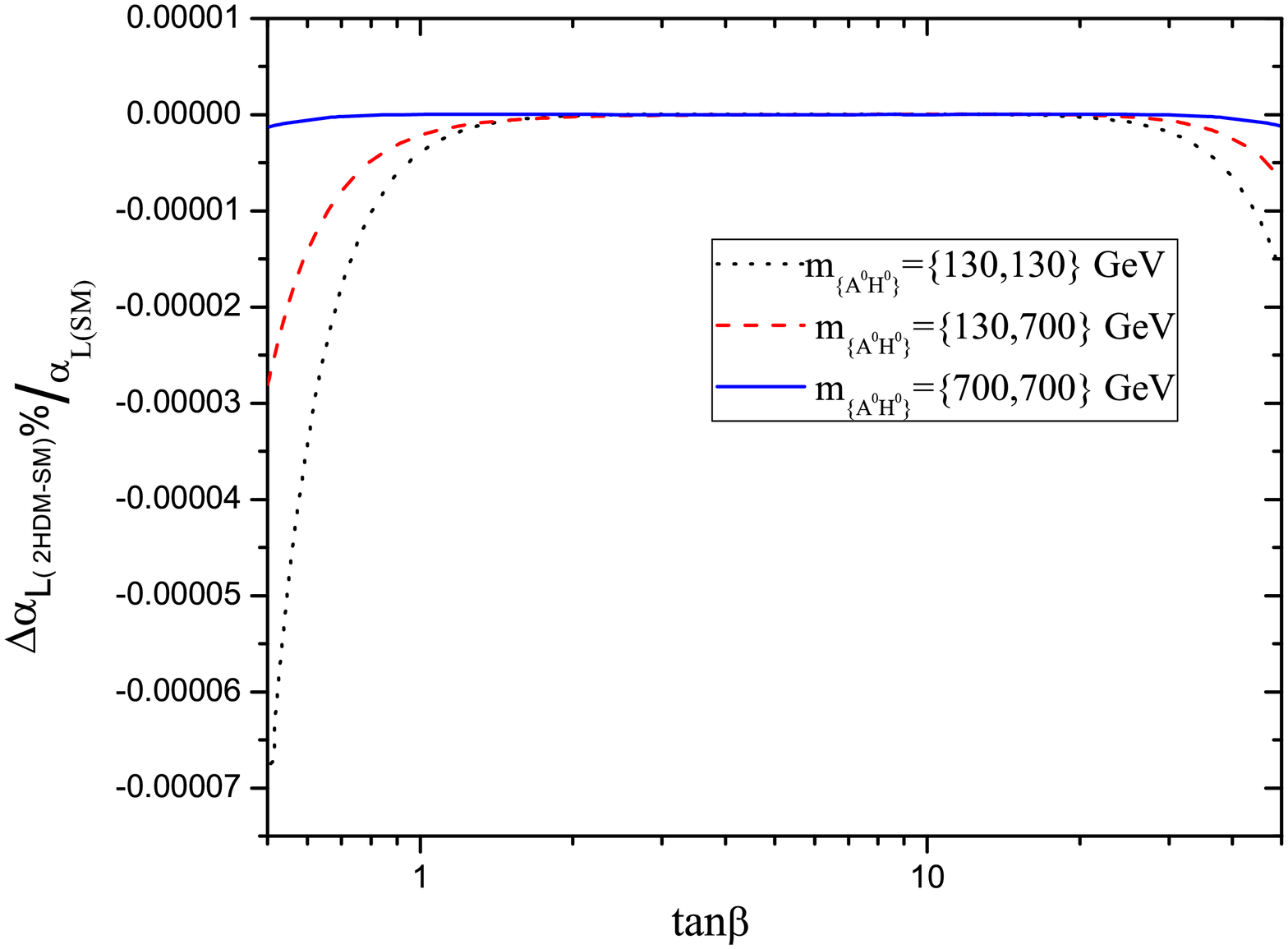}
 \includegraphics[width=7.5cm,height=6cm]{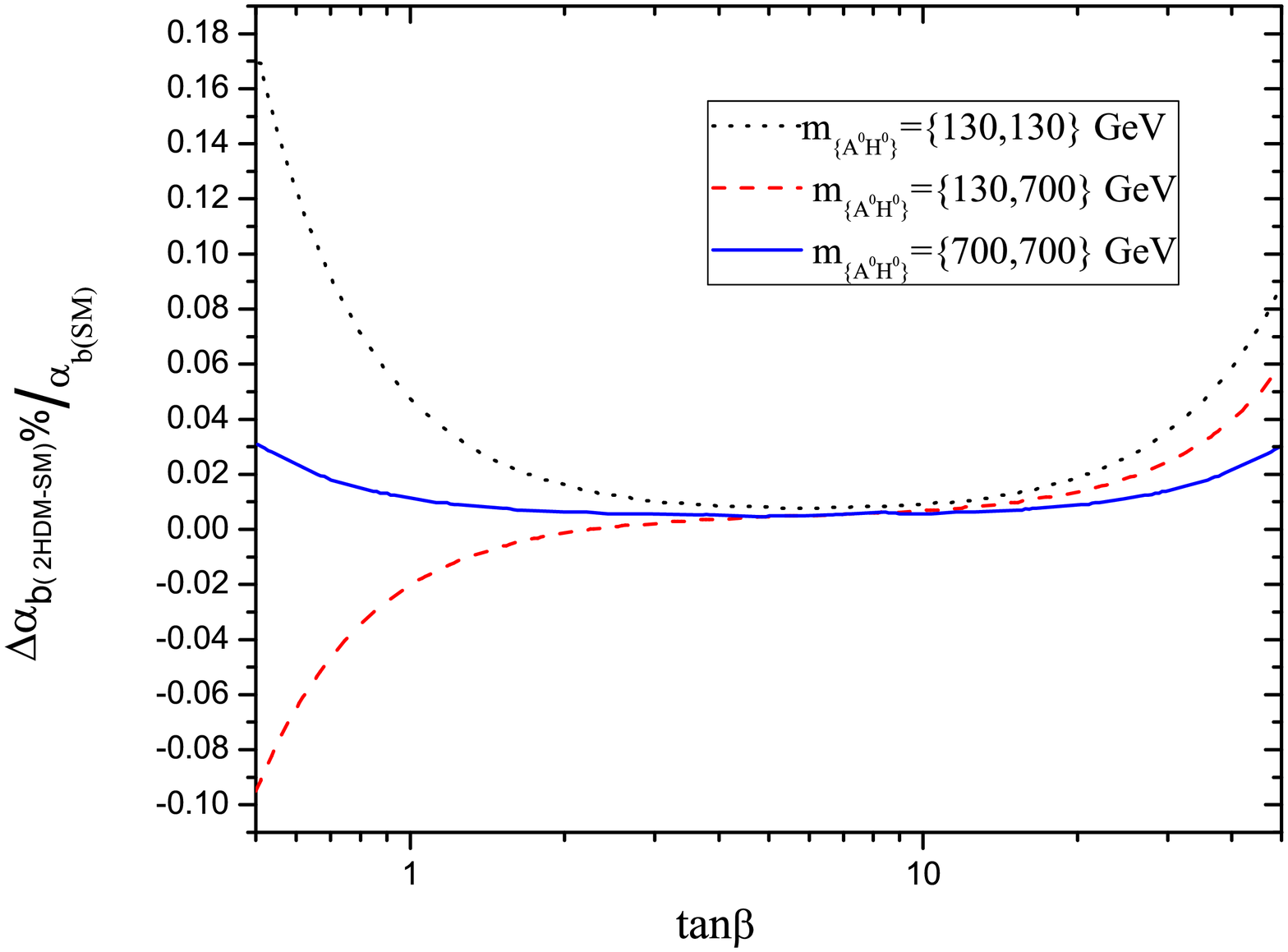} \\
  \caption{The two-Higgs doublet model corrections to the spin analyzing powers as a function of $\tan\beta$.}\label{alphaL2HDM}
\end{figure}

The two-Higgs doublet models are models that extend minimally the Higgs sector of the SM.
One more doublet of complex scalar field is introduced \cite{higgs}. In this work we consider
type II of the two-Higgs doublet model. In type II of the two-Higgs doublet model
one doublet couples to the right-handed down fermions and is
responsible of the down masses; the other doublet couples to the right-handed up fermions
and is responsible of their masses. In addition to two neutral scalar ($h^{0},H^{0}$) and one
pseudo-scalar Higgs bosons ($A^{0}$) the model consists of two charged Higgs bosons ($H^{\pm}$).
Besides the masses of the Higgs bosons, $tan\beta = \frac{v_{1}}{v_{2}}$ (the ratio of the
vacuum expectation values of the two Higgs doubles) and the angle $\alpha$ which describes the
mixing of the two CP-even neutral Higgs bosons are the free parameters of the theory.
The one loop corrections from the model to the $Wtb$ vertex are due the exchange of the Higgs bosons
and the unphysical charged Goldstone bosons. The theoretical and experimental bounds for
free parameters are: $0.1 < tan\beta \lesssim 60,~ m_{H^{\pm}}>320$ GeV/c$^{2}$ and $m_{A^{0},h^{0},H^{0}}>120$ GeV/c$^{2}$
\cite{misiak}.
Figs.~\ref{alphaL2HDM} shows the dependence of the the spin analyzing powers of the $b$-quark and the
charged lepton to the $tan\beta$ while setting $m_{h^{0}} = 120, m_{H^{+}} = 320$ GeV/c$^{2}$ and $\alpha=\beta-\pi/2$.
As it can be seen from Figs.\ref{alphaL2HDM}, within the allowed region of parameters of the
two-Higgs doublet model, the maximum deviation from the
SM values happens for the very small values of $tan\beta$ ($\lesssim 1$). The largest deviation is for the $b$-quark
and is at the order of $0.2\%$. The deviation of $\alpha_{l}$ from its SM value is almost zero.

\subsection{Top-color assisted technicolor (TC2)}

\begin{figure}
  \includegraphics[width=7.5cm,height=6cm]{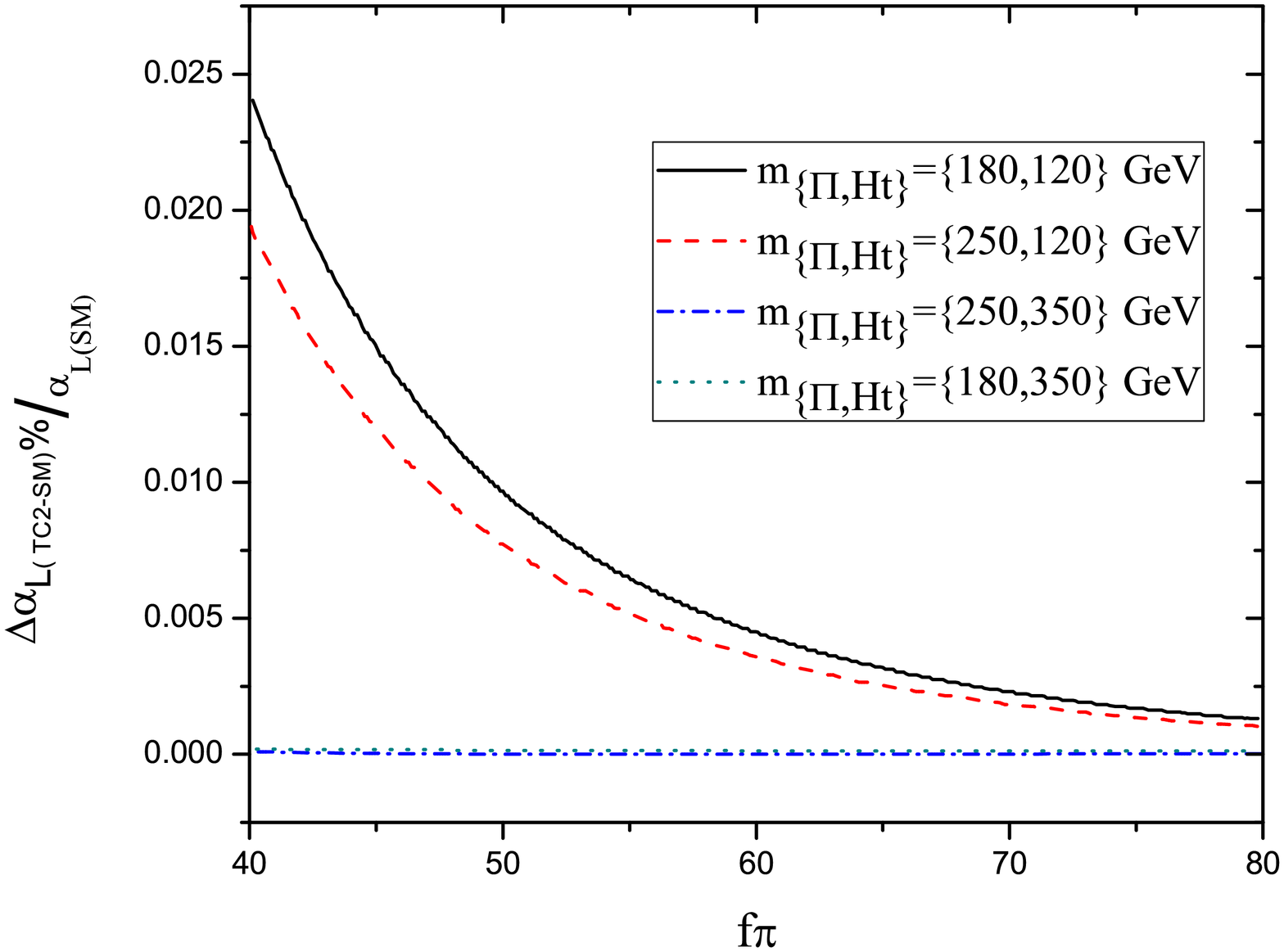}
 \includegraphics[width=7.5cm,height=6cm]{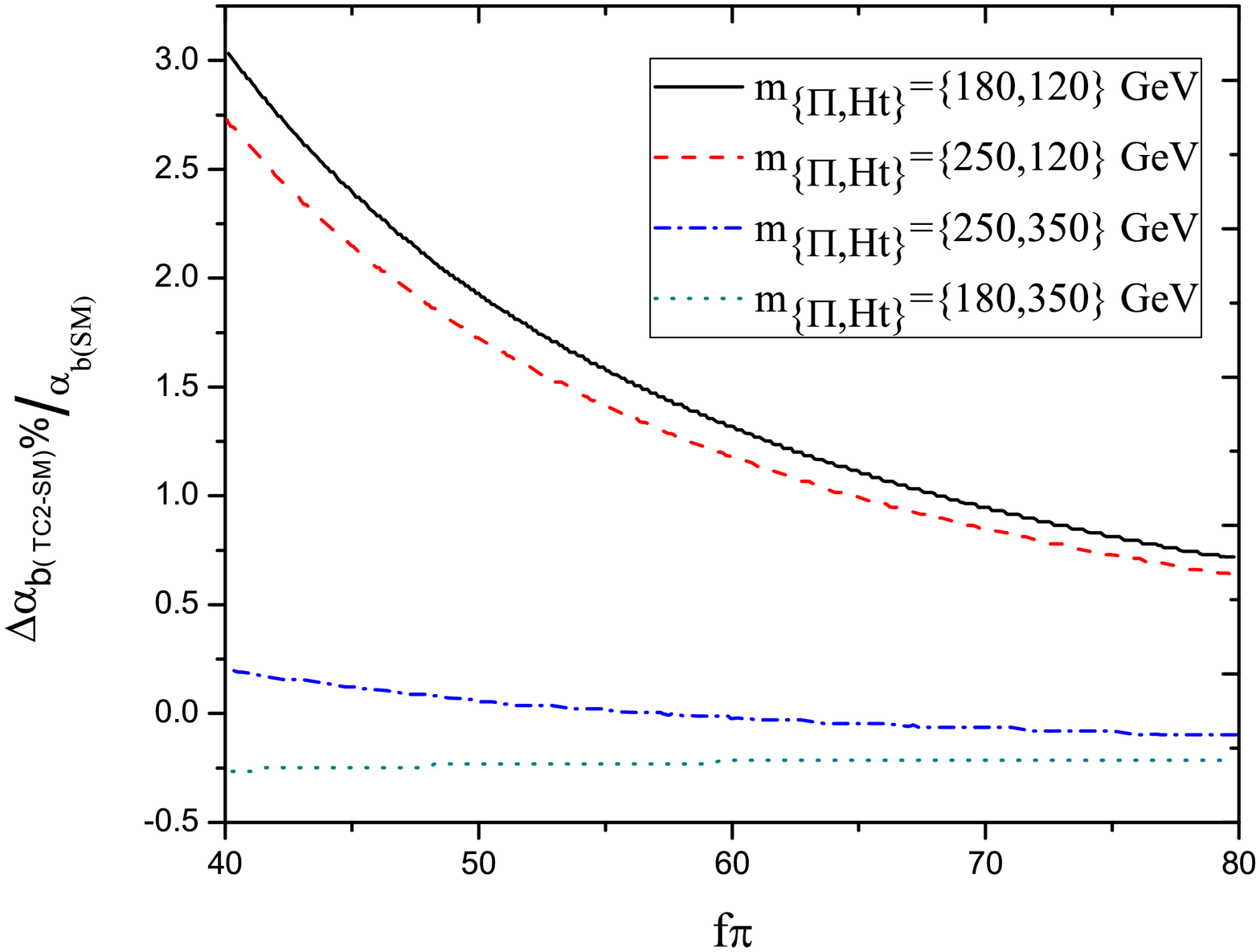} \\
  \caption{The TC2 model corrections to $\alpha_{l,b}$ as a function of $f_{\pi}$.}\label{alphaTC2}
\end{figure}

Top-color assisted technicolor (TC2) offers a dynamical approach for electroweak symmetry breaking (EWSB) \cite{hill}.
TC2 model predicts the existence of three top pions $(\pi^{\pm}_{t},\pi^{0}_{t})$, one
top Higgs $(H^{0}_{t})$ which is a $t\bar{t}$ bound state with large Yukawa couplings to the third family of quarks.
The masses of top pions are expected to be in the range of 180-250 GeV/c$^{2}$ \cite{hill2}.
The masses of new predicted particles and their couplings are the free parameters of the theory.
The one loop corrections from the model to the $Wtb$ vertex are mostly due to the exchange of the top Higgs
$(H^{0}_{t})$ and top pions $(\pi^{\pm}_{t},\pi^{0}_{t})$.
Figs.~\ref{alphaTC2}, show the dependence of the spin analyzing powers as functions of
$f_{\pi}$ for different values of masses of top pions and top Higgs boson within the
allowed ranges of the parameters of the TC2 model.
$f_{\pi}$ is the value of the top quark condensate. The values of other parameters
of the theory are taken the same as what the authors used in \cite{martin11} which are in non-excluded region of the
parameters space of the TC2.

As it could be observed from Figs.\ref{alphaTC2}, the correction of TC2 model to the charged lepton
spin analyzing power is less than $0.02\%$. The deviation of the $b$-quark spin analyzing power
from its SM model value due to the correction from TC2 model is at most $3\%$.

\subsection{Noncommutative extension of the SM}
\begin{figure}
\centering
   \includegraphics[width=8cm,height=5cm]{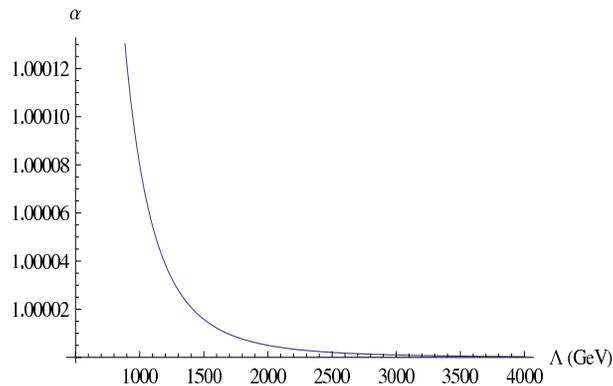} \\
  \caption{The charged lepton spin analyzing power as a function noncommutative characteristic scale.}\label{alphanc}
\end{figure}

The space-time noncommutativity is a possible generalization
of the usual quantum mechanics and quantum field
theory to describe the physics at very short distances of the
order of the Planck length, since the nature of the space-time
changes at these distances \cite{alan}. The noncommutative
spaces can be realized as spaces where coordinate operators
$\hat{x}_{\mu}$ satisfy the commutation relations: $[\hat{x}_{\mu},\hat{x}_{\nu}]=i\theta_{\mu\nu}$
where $\theta_{\mu\nu}$ is a real antisymmetric tensor with the dimension
of $[L]^{2}$. We note that a space-time noncommutativity,
($\theta_{i0} \neq 0$), might lead to some problems with unitarity and
causality. In \cite{moj}, we have calculated the dependence of
the charged lepton spin analyzing power in the top quark decay as
a function noncommutative scale $\Lambda = 1/\sqrt{|\vec{\theta}|}$ where $\vec{\theta}=(\theta_{23},\theta_{31},\theta_{12})$.
Fig.~\ref{alphanc} presents the charged lepton spin analyzing power
versus noncommutativity scale. Within the allowed region of noncommutativity
scale ($\Lambda \gtrsim 1$ TeV) \cite{moj2}, the deviation from the SM prediction is less than $0.01\%$.

Finally, Table \ref{results} shows the deviations which the charged lepton and the $b$-quark
spin analyzing powers can receive from different sources. As it can be seen within the SM
a variation of 2 GeV/c$^2$ of the top quark mass leads to $2.3\%$ change in the $b$-quark spin analyzing power
and no change to $\alpha_{l}$. The maximum deviation with the size of $5\%$ occurs due to the NLO QCD correction for the
bottom quark.

\begin{table}\caption{The deviations which spin analyzing powers can receive from different sources.}
\begin{center}
\begin{tabular}{|c|c|c||c|c|c|}

  \hline
                                 & NLO QCD & $m_{top}\pm 2$ GeV/c$^2$ & 2HDM & TC2 & Noncommutative SM   \\ \hline
   $\Delta\alpha_{l}/\alpha_{l}$ &   $0.1\%$     & 0       & $0.0001\%$ & $0.02\%$ & $0.01\%$  \\ \hline
  $\Delta\alpha_{b}/\alpha_{b}$  &   $5\%$     & $2.3\%$    & $0.2\%$ & $3\%$ & - \\
  \hline
\end{tabular}\label{results}
\end{center}
\end{table}

\section{Conclusions}
The search of the CMS collaboration at the CERN LHC for the $t$-channel
single top quark production is based on the spin correlation of the $t$-channel. The signal extraction
and cross section measurement relies on the angular distribution of the charged
lepton in the top quark decays. In this letter, we investigated the corrections which
the angular distribution of the charged lepton and the $b$-quark receive from the
two-Higgs doublet model, top-assisted technicolor and the noncommutative extension of the
SM. We found that in the frameworks of these models and within the allowed ranges of the parameters
of the models, the spin analyzing powers deviate negligibly from their SM values.
The charged lepton spin analyzing power receive a correction of less than $0.02\%$ in all considered models.
Therefore, even in case of the existence of one of the above models it is safe to use the SM charged lepton
angular distribution to search for the t-channel single top events.

\vspace{0.1cm}
{\bf Acknowledgments}\\
We would like to thank W. Bernreuther, P. Gonzalez, M. Wiebusch for providing us the program
to calculate the anomalous couplings in SM extensions.

\end{document}